\documentclass[prl,amsmath,amssymb,dvips,twocolumn,showpacs]{revtex4}
\usepackage{graphics,epsfig,amssymb}

\def\beq{\begin{equation}} 
\def\eeq{\end{equation}} 
\begin{document}

\title{Quarteting and  spin-aligned proton-neutron pairs in heavy $N=Z$ nuclei}

\author{M. Sambataro$^a$ and N. Sandulescu$^b$}
\affiliation{$^a$Istituto Nazionale di Fisica Nucleare - Sezione di Catania,
Via S. Sofia 64, I-95123 Catania, Italy \\
$^b$National Institute of Physics and Nuclear Engineering, P.O. Box MG-6, 
Magurele, Bucharest, Romania}

\begin{abstract}
We analyze the role of maximally aligned isoscalar pairs  in heavy $N=Z$ nuclei by employing a formalism of quartets. Quartets are superpositions of two neutrons and two protons coupled to total isospin $T=0$ and given $J$. The study 
is focused on the contribution of spin-aligned pairs carrying the angular momentum $J=9$ to the structure of 
$^{96}$Cd and $^{92}$Pd. We show that the role played by the $J=9$ pairs is quite sensitive to the model space 
and, in particular, it decreases considerably by passing from the simple $0g_{9/2}$ space to the more 
complete $1p_{1/2}$,$1p_{3/2}$,$0f_{5/2}$,$0g_{9/2}$ space. In the latter case the description of these nuclei 
in terms of only spin-aligned $J=9$ pairs turns out to be unsatisfactory while an important contribution, particularly in the ground state, is seen to arise
from isovector $J=0$ and isoscalar $J=1$ pairs. Thus, contrary to previous studies, 
we find no compelling evidence of a  spin-aligned pairing phase in $^{92}$Pd. 

\end{abstract}

\maketitle

The role of  spin-aligned isoscalar proton-neutron pairs in the structure of heavy $N=Z$ nuclei is an issue which has received 
much attention in recent years following the first experimental results on the excited states of $^{92}$Pd \cite{cederwall}. 
In Ref. \cite{cederwall} and in an accompanying theoretical analysis \cite{qi} it was suggested that the ground and low-lying 
yrast states of $^{92}$Pd  show evidences of a new spin-aligned pairing phase 
which is fundamentally different from the superfluid phase of  isovector $J=0$ pairs observed in even-even $N\neq Z$ nuclei. 
It was argued, in particular, that the low-lying yrast states of $^{92}$Pd are dominated by the isoscalar $J=9$ pairs and that the approximate equidistance of the yrast states can be interpreted in terms of a simple angular momentum rearrangement of these pairs. This coupling scheme is similar to the stretch pair 
model \cite{danos} and very different from the pair breaking mechanism through which the 
excited states are built in the BCS-like pairing models.

The structure of the yrast states of $^{92}$Pd  has been mainly studied in the framework of the standard shell-model (SM)\cite{qi,isack,coraggio,zamick}. These studies have evidenced the crucial role played by the nuclear interaction in the $J=9$ isoscalar channel in affecting the properties of the low-energy states of this nucleus.  
Consistently, an analogous dominance of isoscalar $J=9$ pairs has been evidenced in the case of $^{96}$Cd \cite{qi}. In this case, large overlaps have been observed  between the SM eigenstates and the corresponding wave functions formulated in terms of isoscalar $J=9$ pairs only \cite{isack,coraggio}. In the case of $^{92}$Pd, instead, the analysis has mostly concentrated on the analysis of the expectation values of the so-called ``pair number operator" \cite{qi}, with $J=9$ pairs exhibiting by far the largest value among isoscalar pairs.
 
The role of the spin-aligned proton-neutron pairs in $^{92}$Pd has also been
studied in the framework of the multi-step shell model \cite{xu}. This approach appears even more appropriate than the standard SM to study the role of proton-neutron pairs in the spectrum of this nucleus since it can be formulated explicitly in terms of these pairs. The conclusions of this analysis,  limited to the case of nucleons confined in the $0g_{9/2}$ orbit, are consistent with those of Refs. \cite{cederwall,qi}.

In this article the role of 
spin-aligned isoscalar pairs in $^{96}$Cd and $^{92}$Pd will be analyzed in a formalism of quartets. We will adopt the same calculation scheme employed in a recent analysis of $sd$ shell nuclei \cite{sasa_a24}. Quartets are defined, in general, as four-body correlated structures characterized by total isospin $T$ and angular momentum $J$. 
Based on the outcome of our analysis of $N=Z$ nuclei in the $sd$ shell, we will introduce only quartets with $T=0$, namely formed by two neutrons and two protons. States of $^{92}$Pd will be described as superpositions of products of two quartets coupled to given $J$.
The advantage of this calculation scheme, which
conserves all symmetries and gives results of an accuracy comparable to that of
SM calculations \cite{sasa_a24,sasat1,sasaj0},  is a simple structure of the wave function, 
well-adapted for investigating the underline correlations. 
We will carry out calculations with quartets in their most complete form and verify that the spectrum so generated provides a 
satisfactory description of the experimental spectrum of $^{92}$Pd. Then we will explore the validity of various approximations based on quartets built only by some selected types of pairs. From an analysis of the resulting spectra and electromagnetic transitions as well as of the overlaps among wave functions in the various approximations 
we will extract information on the structure of the low-lying states of $^{96}$Cd and $^{92}$Pd.

Qartets are defined as \cite{sasa_a24}
\begin{eqnarray}
Q^+_{\alpha ,JM,TT_z}&=&\sum_{i_1j_1J_1T_1}\sum_{i_2j_2J_2T_2}
C^{(\alpha )}_{i_1j_1J_1T_1,i_2j_2J_2T_2}\nonumber\\
&&\times\Bigl[ [a^+_{i_1}a^+_{j_1}]^{J_1T_1}[a^+_{i_2}a^+_{j_2}]^{J_2T_2}\Bigr]^{JT}_{MT_z},
\end{eqnarray}
where $J(T)$ and $M(T_z)$ are, respectively, the total angular momentum (isospin) and the relative projections. 
The indices $i$ and $j$ denote the quantum numbers of the single-particle states considered in the calculations.
We work in a spherical single-particle basis and therefore, according to the standard SM notation,  $i\equiv \{n_i,l_i,j_i \}$.

The collective quartet (1) provides the exact SM wave function of a system with four active nucleons outside a closed core. Values of the isospin $T$ range in the interval (0,2) and, depending upon the projection $T_z$, all possible combinations of protons and neutrons can be represented. 
Systems with eight active nucleons outside an inert core can be described in a basis
formed by the tensorial product of two quartets (1), i.e.,
\beq
[Q^+_{\alpha_1,J',T'}  \otimes Q^+_{\alpha_2,J'',T''}]^{J,T},
 \eeq
where $J,T$ are the spin and the isospin of the calculated state. If all possible quartets which can be formed within a given model space are inserted in (2), this basis spans the entire Hilbert space and the corresponding spectrum is exact.
Since the basis (2) is overcomplete, an exact calculation in this framework would be more difficult than in standard SM. However, if a satisfactory approximation of the exact spectrum is obtained in terms of only a limited set of quartets, this can give us an insight into
the relevant degrees of freedom of the eigenstates. This approach can in principle be extended to any system with $4n$ active nucleons and it will be referred in the following as quartet model (QM).

The  calculation scheme described above will be applied in this work to $^{96}$Cd and $^{92}$Pd.
By assuming $^{100}$Sn as the inert core of reference, $^{96}$Cd is a system with two proton holes and two neutron holes in this core and it will be therefore described as a single quartet. As already noticed, in this case the SM and QM approaches coincide. $^{92}$Pd has instead four proton holes and four neutron holes with respect to  $^{100}$Sn and it will be therefore described as a superposition of two-quartet states of the type shown in Eq. (2).
Two basic problems are encountered in this case: which quartets to involve in the calculations and how to construct these quartets. Here
we  adopt a {\it static} formulation of the quartets,  which means that as quartets defining the basis (2) we assume those describing
the low-lying states of $^{96}$Cd. More precisely, we will employ the quartets associated  with the
positive-parity yrast states of $^{96}$Cd up to $J=8$. These are all $T=0$ quartets. 

The structure of nuclei with mass number immediately below $A=100$ is expected to be dominated by the  $0g_{9/2}$ orbit and, in some studies, calculations have been restricted to this orbit only \cite{isack,coraggio,xu}. To understand better the structure of these nuclei, in this work
we will perform calculations within three different model spaces. These are composed by the  orbits $(0g_{9/2})$,  $(1p_{1/2}0g_{9/2})$ and $(1p_{3/2}0f_{5/2}1p_{1/2}0g_{9/2})$. We will refer to them as $g$, $pg$ and
$fpg$ spaces, respectively. The latter will also be referred to as the full (model) space. The interactions that we will use are
a renormalized SLGT0  for the $g$ space \cite{isack}, 
the F-FIT by Johnstone and Skouras \cite{skouras} for the $pg$ space and the JUN45 \cite{june45} for the $fpg$ space.

In order to investigate the contribution of various pairs of nucleons to the physical states, in addition to QM calculations performed with full quartets, namely quartets receiving contributions from all possible pairs 
(hereafter we will refer to these calculations simply as QM calculations), we will perform approximate QM calculations in which only selected types of pairs will take part in the formation of the quartets (1). In particular, we will discuss:\\
i) the QM$_{SA}$ scheme, in which the quartets (1) are built in terms of only isoscalar $J=9$ pairs and therefore carry only the component
\beq
\Bigl[ [a^+_{i_1}a^+_{j_1}]^{J_1=9 T_1=0}[a^+_{i_2}a^+_{j_2}]^{J_2=9 T_2=0}\Bigr]^{JT=0};
\eeq
ii) the QM$_{IV}$ scheme, in which quartets are superpositions of the non-collective quartets
\beq
\Bigl[ [a^+_{i_1}a^+_{j_1}]^{J_1=0 T_1=1}[a^+_{i_2}a^+_{j_2}]^{J_2=J T_2=1}\Bigr]^{JT=0},
\eeq
formed only with isovector pairs (notice that one of the two pairs is constrained to have $J=0$ while the other one is responsible for the angular momentum of the quartet);\\
iii) the QM$_{R}$ scheme, in which quartets carry both the components (3) and (4);\\
iv) the QM$_{J1}$ scheme, in which quartets have the component
\beq
\Bigl[ [a^+_{i_1}a^+_{j_1}]^{J_1=1 T_1=0}[a^+_{i_2}a^+_{j_2}]^{J_2=1 T_2=0}\Bigr]^{JT=0},
\eeq
formed with isoscalar $J=1$ pairs only, in addition to the isovector-type component (4).
We observe that the QM$_{IV}$ and QM$_{J1}$ schemes become identical when the $J$ of the quartet is larger than 2.

\begin{figure}
\begin{center}
\includegraphics[width=3.5in,height=5.5in,angle=0]{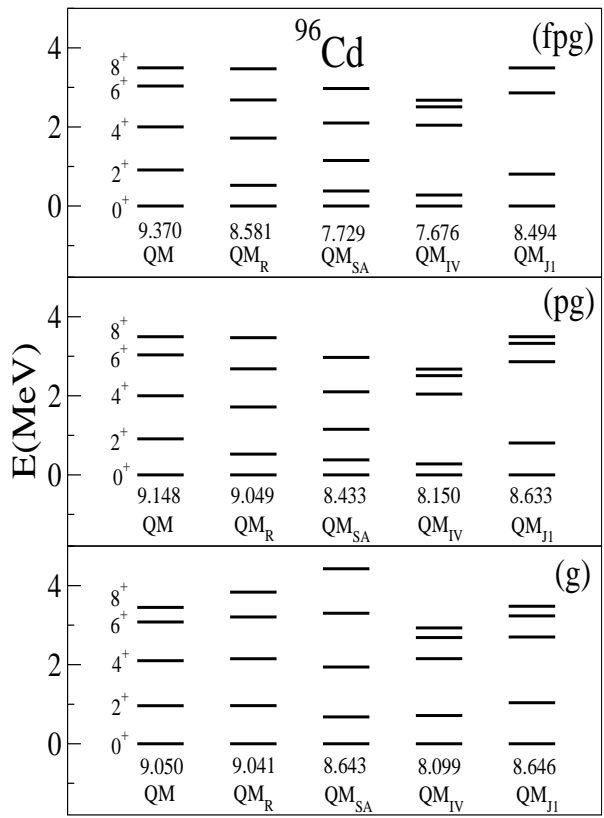}
\caption{
Low-energy yrast spectra of $^{96}$Cd obtained in the quartet model (QM) and in the
various approximations  QM$_i$ explained in the text. From bottom to top, the three panels correspond to calculations done within the model spaces $g$, $pg$ and $fpg$, respectively.
The number below each spectrum gives the ground state correlation energy, namely the difference 
between the total ground state energy and the energy in the absence of interaction.}

\end{center}
\end{figure}

Fig. 1 shows the spectra that are obtained for $^{96}$Cd within the just defined approximation schemes and for the three model spaces  $g$, $pg$ and
$fpg$. A number of things are worthy being noticed. The QM$_{SA}$ scheme shows an increasing difficulty in reproducing the QM spectrum with increasing the size of the model space.
In the $fpg$ space, the QM$_{SA}$ spectrum is quite compressed and the ground state correlation energy (defined as the difference between the ground state energy and the energy of this state in the absence of interaction) is far from the exact value. Even in the most favorable case (the $g$ space), however, one can still observe significant deviations in the ground state correlation energy (which is underestimated by about 400 KeV) as well as in the energies of the $2^+$ and $8^+$ 
states.  A considerable improvement of the spectrum, both in the $g$ and $pg$ spaces, is obtained in the QM$_R$ approximation which mixes  the spin-aligned and seniority coupling schemes. When passing to the $fpg$ space, however,
even the QM$_{R}$ scheme appears to be no longer fully adequate. In this case we have verified that  in order to restore a good agreement with the QM spectrum it is sufficient to add to the QM$_R$ approximation  the contribution of the quartets built by two pairs with $J=2$.

\begin{figure}
\begin{center}
\includegraphics[width=3.5in,height=5.5in,angle=0]{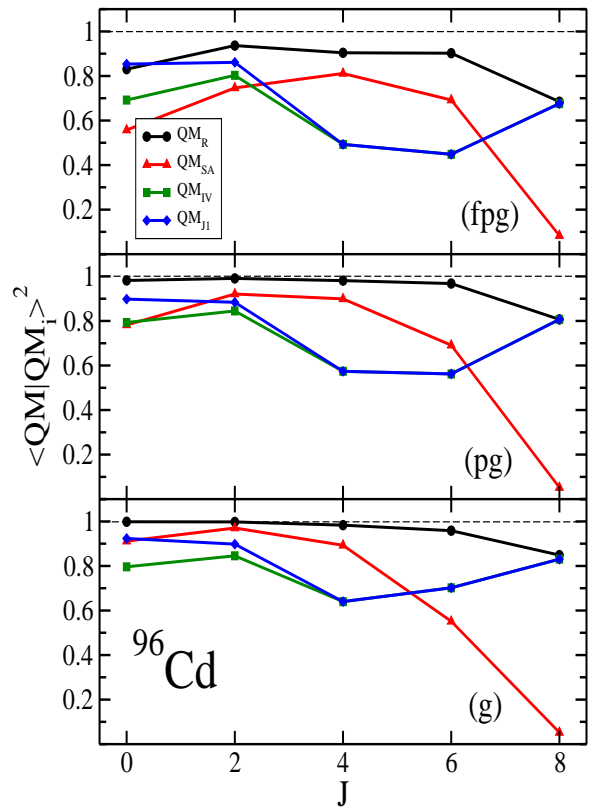}
\caption{
(Color online) Overlaps  between the QM low-lying yrast states of $^{96}$Cd and the corresponding eigenstates  in the various QM$_i$ approximations explained in the text. From bottom to top,  the three panels correspond to calculations done within the model spaces $g$, $pg$ and $fpg$, respectively. }
\end{center}
\end{figure}

Fig. 2  shows the square of the overlaps $\langle$QM$|$QM$_i\rangle$, where QM$_i$ are the eingenfunctions of $^{96}$Cd
corresponding to the  approximate schemes defined above. This figure clearly shows how 
the contribution  of the spin-aligned $J=9$ pairs to the physical states evolves by varying the model space. In particular, for the ground state one
notices that  the squared overlap $\langle$QM$|$QM$_{SA}\rangle^2$ decreases from 0.91 to 0.56 when one goes from $g$ to  $fpg$ model 
space. It is also interesting to observe that, in the full space,
the QM$_R$ and QM$_{J1}$ schemes generate ground states which have almost the same overlaps with the exact ground state and, in addition,
predict binding energies which are close to each other (Fig. 1). Consequently, in this case the role of isoscalar pairs with $J=1$ is comparable with 
that of the spin-aligned $J=9$ pairs. For all other eigenstates and model spaces the QM$_R$ scheme is the one which gives the largest overlaps with the exact wave functions. As a final remark relative to Fig. 2, we observe the negligible role of the spin-aligned pairs in the $J=8$ yrast level which is instead much better represented in the QM$_{IV}$ scheme.

\begin{figure}
\begin{center}
\includegraphics[width=3.5in,height=5.5in,angle=0]{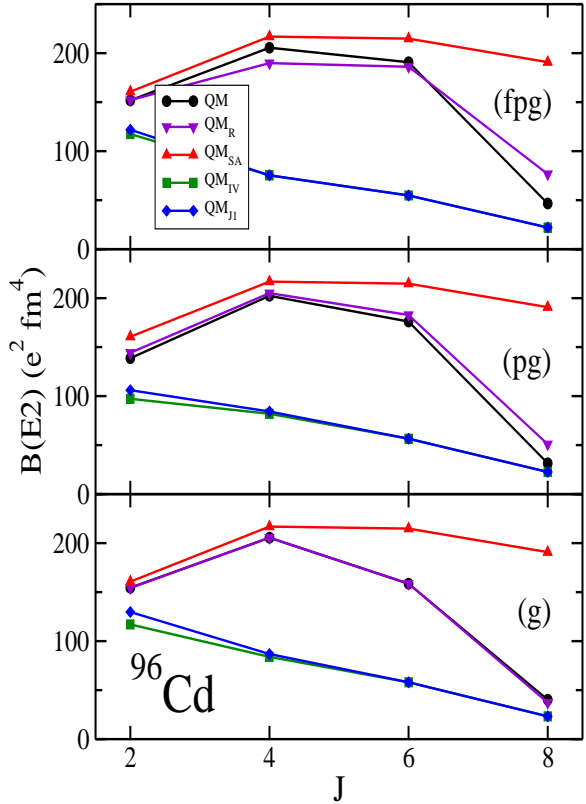}
\caption{
(Color online) $B(E2; J \rightarrow J-2)$  values between the low-lying yrast levels of $^{96}$Cd 
in the quartet model (QM) and in the various QM$_i$ approximations
explained in the text. From bottom to top,  the three panels correspond to calculations done within the model spaces $g$, $pg$ and $fpg$, respectively.}
\end{center}
\end{figure}

Fig. 3 shows the $B(E2; J \rightarrow J-2)$ values between the yrast states of $^{96}$Cd in the various approximation schemes. For the $E2$ operator we have adopted standard values of the effective charges $e_p=1.5e$ and $e_n=0.5e$ and of the (squared) oscillator length $b^2\approx 41.4/\hbar\omega$ fm$^2$,
$\hbar\omega =45A^{-1/3}-25A^{-2/3}$.
Similar to what observed in the analysis of the energies and the overlaps, the QM$_{R}$ scheme is the one which gives the best overall fit of the exact values among the various approximation schemes. 
Differently from the case of the energies and the overlaps, however, the QM$_{SA}$ results for the $B(E2)$'s do not exhibit the rapid deterioration with increasing the size of the model space which was instead observed in Figs. 1 and 2.

\begin{figure}
\begin{center}
\includegraphics[width=3.5in,height=5.5in,angle=0]{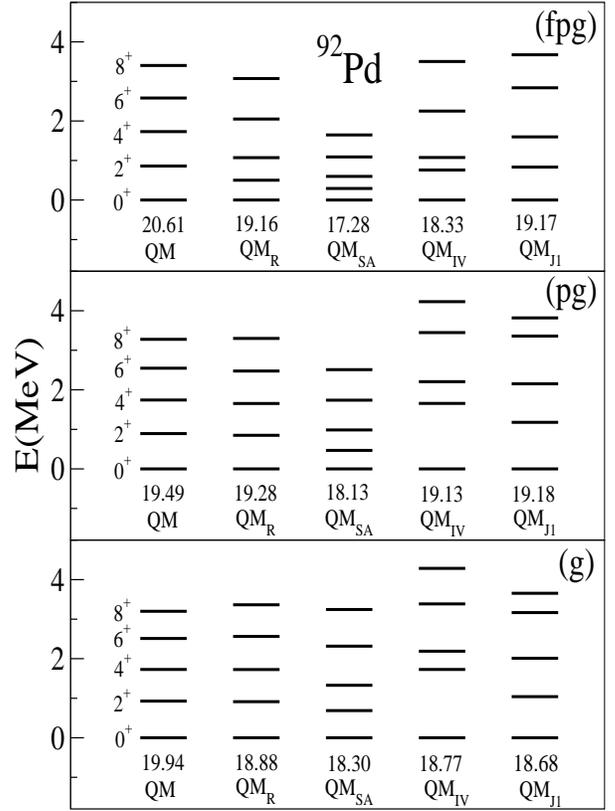}
\caption{
Low-energy yrast spectra of $^{92}$Pd obtained in the quartet model (QM) and in the
various approximations  QM$_i$ explained in the text. From bottom to top, the three panels correspond to calculations done within the model spaces $g$, $pg$ and $fpg$, respectively.
The number below each spectrum gives the ground state correlation energy, namely the difference 
between the total ground state energy and the energy in the absence of interaction.}

\end{center}
\end{figure}

As anticipated, the analysis of $^{92}$Pd is based on the same quartets already employed for $^{96}$Cd. These quartets are used to construct the basis (2) and the spectrum of $^{92}$Pd is generated by diagonalizing the Hamiltonian in this basis. This spectrum (limited to the low-lying yrast states only) is shown in Fig. 4 for the various approximations. The notation is the same adopted for $^{96}$Cd. Only three excited levels are known experimentally and their energies are (in MeV): E(2$^+$)=0.874, E(4$^+$)=1.786, E(6$^+$)=2.536 \cite{cederwall}. The QM spectra reproduce well these states in all model spaces.
It is worthy noticing that the QM ground state turns out to be basically composed by $J=0$ quartets only since we have verified that, in all three model spaces, a state product of two such quartets accounts by itself for more than 99$\%$ of the ground state correlation energy.

\begin{figure}
\begin{center}
\includegraphics[width=3.5in,height=5.5in,angle=0]{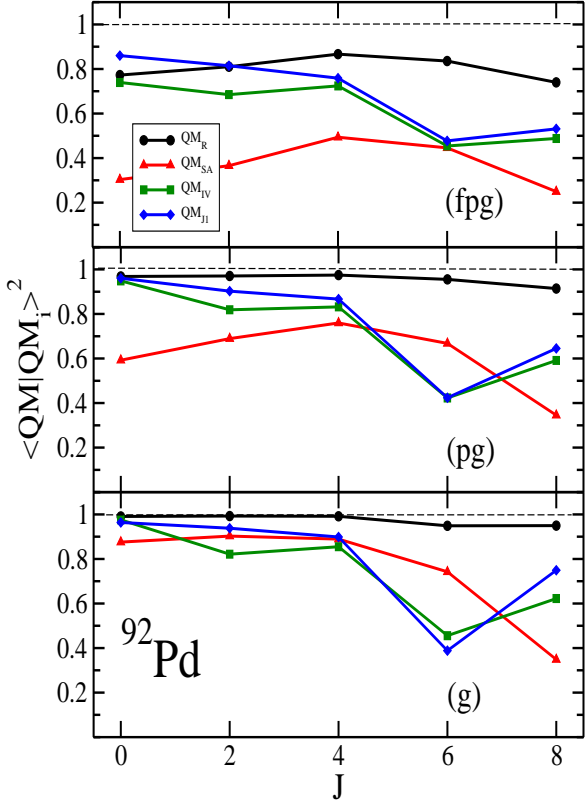}
\caption{
(Color online) Overlaps  between the $QM$ low-lying yrast states of $^{92}$Pd and the corresponding eigenstates  in the various QM$_i$ approximations explained in the text. From bottom to top,  the three panels correspond to calculations done within the model spaces $g$, $pg$ and $fpg$, respectively. }
\end{center}
\end{figure}

Fig. 4 has several features in common with Fig. 1. The evolution of the QM$_{SA}$ spectrum when passing from the  $g$ space
to the full space looks quite similar in the two figures.  In Fig.4, however, one notices that the mismatch 
between  the QM and QM$_{SA}$ spectra has become even more pronounced than in $^{96}$Cd in the calculation relative to the $fpg$ space.
Still in Fig. 4 one sees  that  the QM$_R$ scheme generates a good spectrum in the $g$ and $pg$ spaces while it is not fully adequate in the $fpg$ space.  We have verified that adding to the QM$_R$ quartets the quartets 
built by two $J=2$ pairs, as we have done for $^{96}$Cd, one  gets also in the full space a good agreement 
with the SM spectrum.

\begin{figure}
\begin{center}
\includegraphics[width=3.5in,height=5.5in,angle=0]{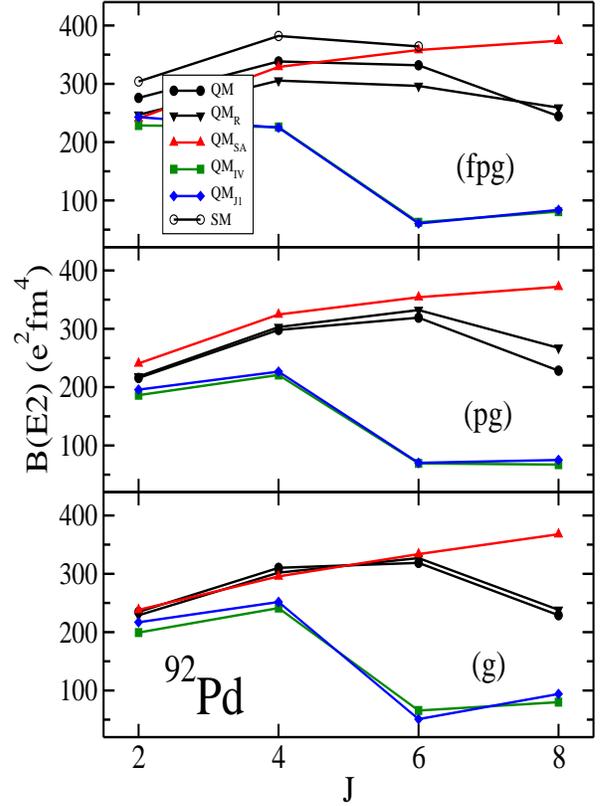}
\caption{
(Color online) $B(E2; J \rightarrow J-2)$  values between the low-lying yrast levels of 
$^{92}$Pd in the quartet model (QM) and in the various QM$_i$ approximations
explained in the text. From bottom to top,  the three panels correspond to calculations done within the model spaces $g$, $pg$ and $fpg$, respectively.}
\end{center}
\end{figure}

As seen in Fig. 4, in the $fpg$ space the QM$_{J1}$ scheme generates a ground state whose energy is 
basically identical to that of the QM$_R$ scheme. Moreover, one can observe that the low-lying states 
generated by QM$_{J1}$ are even closer to the QM results than the ones predicted by QM$_R$. 
This confirms the outcome of the analysis in $^{96}$Cd on the relevant role of the isoscalar $J=1$ pairs 
in the full model space calculations. 

A further evidence of the role of the $J=1$ pairs arises from the observation of the overlaps $\langle$QM$|$QM$_i\rangle$, the squares of which are shown in Fig. 5. In the ground state of the $fpg$ model space, the overlap in the QM$_{J1}$ scheme is seen to be the largest one among those shown in the figure. With only this exception, the QM$_R$ scheme gives the best results in the full space while, on the contrary, QM$_{SA}$ generates by far the worst results.
These facts suggest that the isovector 
component (4) of the QM$_R$ quartets plays a leading role over the spin-aligned part (3). 
As a  confirmation of that, we see that the overlaps $\langle$QM$|$QM$_{IV}\rangle$ are considerably 
larger than the $\langle$QM$|$QM$_{SA}\rangle$ ones (with the only exception of the $J=6$ state). 
Altogether these results do not show any dominance of the spin-aligned $J=9$ pairs in the 
low-lying yrast states of $^{92}$Pd. What emerges is instead the relevant 
role played by the isovector  $J=0$ pairs and isoscalar $J=1$  pairs in the structure of 
the  ground state (we remind that the ground state of $^{92}$Pd is to a very large extent
a product of  two $J=0$ quartets and therefore the components (4) of the QM$_{IV}$ quartets 
in this state are formed exclusively by $J=0$ pairs). These conclusions remain valid 
also in the $g$ and $pg$ spaces although one can notice that the differences among the various 
approximation schemes become less apparent with reducing the model space. 
The increased role of $J=0$ and $J=1$ pairs 
relative to $J=9$ pairs that is observed in the calculations done in the full space is plausibly related to the fact that the former pairs get contributions from nucleons sitting in all the orbitals of the $fpg$ space while the spin-aligned $J=9$ pairs can be formed only in the orbital $0g_{9/2}$.  
        
In Fig. 6,  we display  the $B(E2; J \rightarrow J-2)$ values for the transitions  between the yrast states of $^{92}$Pd . In the case of the full space, we also include the results of SM calculations \cite{zuker}. The QM values exhibit the same trend as the SM results although with a modest underestimation. As we have verified, this can be reduced by adding extra quartets in the QM basis. Although not exhibiting a deterioration with increasing the size of the model space comparable with that of the spectrum (as for $^{96}$Cd), the QM$_{SA}$ values in the full space are characterized by a trend which deviates from that of SM and QM results. In particular, at variance with SM and QM results, one can observe that in the full space 
the QM$_{SA}$ predicts an  increase of $B(E2)$ values from $J=4$ to $J=8$. Large deviations from the QM 
values are also observed in all spaces for the QM$_{IV}$ and QM$_{J1}$ results for $J\geq 2$ and for the QM$_{SA}$ results for $J=8$. In all cases these deviations are consistent with the trend of the overlaps in Fig. 5.
  
By summarizing,  in this paper  we have studied the role played by  the spin-aligned $J=9$ proton-neutron pairs on the structure of  $^{96}$Cd and $^{92}$Pd. 
The analysis has been carried out in the framework of a quartet model.
We have found that the contribution of spin-aligned $J=9$ pairs to the structure of low-lying states of these nuclei is strongly depending on the
model space and it decreases considerably passing from the simple $(0g_{9/2})$ space to the more complete $(1p_{3/2}0f_{5/2}1p_{1/2}0g_{9/2})$ space. 
In the ground state of $^{92}$Pd, in particular, the role of isoscalar $J=1$ and isovector $J=0$ pairs has been found prominent with respect to that of isoscalar $J=9$ pairs.

\vskip 0.3cm

{\it Acknowledgments} 
We thank Danilo Gambacurta for useful discussions. 
One of us (N.S.) thanks also the nuclear physics group of the Royal Institute of Technology, Stockholm, for many valuable discussions on the topics studied in this paper. 
This work was supported by the Romanian Ministry of Education and Research
through the grant Idei nr 57.

\end{document}